\documentclass{aa}  
\usepackage{graphicx}
\usepackage{txfonts}
\usepackage{natbib}
\bibpunct{(}{)}{;}{a}{}{,}
\begin{document}
%
   \title{Quantifying the contamination by old main-sequence stars in 
   young moving groups: the case of the Local Association
   \thanks{On-line tables with the data are available in the CDS.}}

  \titlerunning{Contamination by old main-sequence stars in 
                         young moving groups
                         }

   \subtitle{}

   \author{J. L\'opez-Santiago\inst{1, 2}
          \and
          G. Micela\inst{2}
          \and
          D. Montes\inst{1}
          }

   \offprints{J. L\'opez-Santiago}

   \institute{Departamento de Astrof\'{\i}sica y Ciencias de la Atm\'osfera,
              Universidad Complutense de Madrid,
              E-28040 Madrid, Spain\\
              \email{jls@astrax.fis.ucm.es; dmg@astrax.fis.ucm.es}
          \and
              INAF - Osservatorio Astronomico di Palermo Giuseppe S. Vaiana,
              Piazza Parlamento 1, I-90134 Palermo, Italy\\
              \email{giusi@astropa.unipa.it}
             }

   \date{Received ; accepted }

 
  \abstract
   {The associations and moving groups of young stars are excellent 
    laboratories for investigating stellar formation in the solar 
    neighborhood. 
    Previous results have confirmed that a non-negligible fraction of old 
    main-sequence 
    stars is present in the lists of possible members of young stellar kinematic
    groups. A detailed study of the properties of these samples is needed
    to separate the young stars from old main-sequence stars with similar 
    space motion, and identify the origin of these structures.
    }
   {Our intention is to characterize members of 
    the young moving groups, 
    determine their age distribution, and 
    quantify the contamination by old main-sequence 
    stars, in particular, for the Local Association.
    }
   {We used stars possible members of the young
    ($\sim 10 - 650$ Myr) moving groups from the literature. 
    To determine the age of the stars, we used several suitable age
    indicators for young main sequence stars, i.e.,  
    X-ray fluxes 
    from the Rosat All-sky Survey database, 
    photometric data 
    from the Tycho-2, Hipparcos, and 2MASS database.
    We also used spectroscopic data, in particular the equivalent width 
    of the lithium line Li~\textsc{i} $\lambda$6707.8 \AA\ and H$_\alpha$,
    to 
    constrain the range of ages of the stars.
    }
   {By combining photometric and spectroscopic data, we were able to 
   separate the young stars ($10 - 650$~Myr) from the old ($> 1Gyr$) field ones. 
   We found, in particular, that the Local Association is contaminated 
   by old field stars  at the level of $\sim 30\%$. 
   This value must be considered as the contamination for our 
   particular sample, and not of the entire Local Association.
   For other young 
   moving groups, it is more difficult to estimate the fraction of old stars
   among possible members. However, the level of X-ray emission
   can, at least, help to separate two age populations: stars with 
   $< 200$~Myr and stars older than this.
   }
   {Among the candidate members of the classical moving groups, there is a 
   non-negligible fraction of old field stars that should be taken into
   account when studying
   the stellar birthrate in the solar neighborhood. Our results are consistent 
   with a 
   scenario in which the moving groups contain both groups of young stars 
   formed in a recent star-formation episode and old field stars with similar 
   space motion. Only by combining  
   X-ray and optical spectroscopic data is it possible to distinguish between 
   these two age populations. 
   }


   \maketitle
%

\section{Introduction}
\label{s:intro}

It is well known that star formation takes place inside giant molecular clouds
and, indeed, 
stellar class I and class II objects (classical T\,Tauri stars, cTTS) 
are found mainly in these regions. Young stars 
are strong X-ray emitters. In particular, cTTS and weak-line T\,Tauri stars
(wTTS) have emission levels above those observed for 
main-sequence (MS) stars. Thus, X-ray surveys should 
preferentially detect cTTS and wTTS inside the forming regions. 
In a study of the spatial distribution of 
X-ray coronal emitters detected with \textit{ROSAT}, \citet{gui98} found 
different over density regions of X-ray active stars coinciding with the 
position of the nearby molecular complexes: Chamaleontis I and II 
($d \sim 100$\,pc), Taurus-Auriga ($d \sim 140$\,pc), Scorpius-Centaurus-Lupus 
($d \sim 150$\,pc), and Ophiuchus ($d \sim 160$\,pc). 
In contrast, other regions exhibited the average Galactic plane characteristics. 
It is logical to expect post-T\,Tauri stars to be found 
in the proximity of these stellar complexes, but not many 
post-T\,Tauris are observed near the molecular clouds \citep{her78}. 
\citet{mam01} pointed out that the main process by which stars are dispersed, 
i.e., the evaporation of stellar clusters and/or stellar complexes, is a 
slow process, but high dispersal velocities are required to transport
stars some tens of parsecs from their parental clouds on short timescales.
A dispersion velocity of $1-2$ km\,s$^{-1}$
is sufficient to separate a star from its formation locus by as much as 10\,pc in 
$5-10$ Myr, and, thus, many young stars may travel to considerable distances
from their parental clouds.
Since many of the nearby star-forming regions are situated in the southern 
hemisphere, many young PMS stars with low declinations
are found in the solar neighborhood \citep{tor06}. 

A series of associations of late-type stars with ages ranging
from 8 to 50 Myr and having similar space motion have been discovered in our
neighborhood: TW~Hya, $\beta$~Pic, AB~Dor, $\eta$ and $\epsilon$~Cha, 
Octans, and Argus associations, and the Great Austral complex (GAYA), which 
includes the Tucana-Horologium, Columba, and Carina associations
\citep{zuc04,tor08}.
A detailed study of the space motion of these young associations has shown 
that many of them were close to a molecular cloud in the past, in particular
the Scorpius-Centaurus-Lupus complex \citep{zuc04} and the dense clouds 
in Ophiuchus and Corona Australis \citep{mak07,fer08}. Thus, a significant 
portion of the young stellar population in the solar vicinity is a remnant 
of the star formation that took place in the past in the region of the 
Galaxy occupied by Sco-Cen-Lup. 

Slightly older groups of stars ($\sim 50 - 650$ Myr) with similar space motion have
also been detected in the solar neighborhood. They are the classical stellar
kinematic groups, or moving groups \citep[see][and references therein]{mon01a,
lop06}. In contrast to the young stellar associations, 
the stars belonging to a moving group are situated all over
the sky. The dispersion velocity and the differential Galactic rotation, 
acting together over millions of years, have caused the wide separation
of their members. A large spread in the velocity space of these stars is also 
observed \citep{sku99}. 
%
The idea of moving groups consisting of coeval stars is rather 
controversial.
The over density of stars in some regions of the $UV$-plane could also be
the result of dynamical perturbations caused by spiral waves
\citep[e.g.][]{fam05}. However, several works showed that
different age subgroups are situated in the same region of the Galactic velocity 
plane as the classical moving groups \citep[see][]{asi99}. 
\citet{fam07} studied a large sample of stars sharing the space 
motion of the Hyades cluster, and determined that part of these stars were 
surely associated to it in the past, while the remainder are
older stars trapped at resonance. Together with the result afore mentioned
this suggests that those regions of the $UV$-plane consist of both field-like 
stars and young coeval ones \citep{fam07,fam08,ant08,kle08,fra09,zha09}.
This problem is not exclusive to moving groups, but also to 
stars in young T associations. Thus, \citet{ber06} found that 
among the stars with similar proper motions in the Taurus region, 
there are field stars and pre-main-sequence ones. 
These authors also noticed that it is impossible to distinguish 
kinematically between pre-main sequence and field stars in their sample
and that it is crucial to remove possible interlopers before searching 
for a moving group of young stars. In particular, 
distinguishing between members --\,i.e., coeval stars\,-- of the moving 
groups and other field-like stars with similar space motion is necessary
to investigate properly their contribution to the stellar population 
of the solar neighborhood.
 

Studies of the stellar content in flux-limited shallow X-ray surveys 
\citep{fav93, mic07, lop07} have detected an excess of yellow stars in 
observations that cannot be reproduced by standard galactic models using any 
form of continuous star-formation rate. A possible explanation in terms of binary 
systems with a yellow primary star and a secondary M dwarf, which responsible 
for the X-ray emission, was proposed by \citet{mic07}. 
However, this does not explain the apparent excess of stars with high X-ray 
fluxes detected in the $\log N - \log S$ diagram \citep{lop07}. This group of
high X-ray emitters also have low scale height, which is typical of young
stars. The optical follow-up of the coronal sources in the 
\textit{Einstein} Medium Sensitivity Survey \citep{sci95} demonstrated 
that many of them are indeed young lithium-rich stars. A similar result
was obtained for the North Ecliptic Pole (NEP) survey \citep{aff08}.
It agrees with a 
scenario in which the solar neighborhood consists of a standard population
of stars formed with a constant star-formation rate, and an additional young stellar 
population. In this case, we should be able to detect this young population
from their X-ray emission, since young stars are high X-ray emitters.
A different question is how to explain their presence in the solar vicinity 
and their origin. 

In this work, we investigate the contamination by old main-sequence 
stars in samples of possible members of the young moving groups.
Our main goal is to quantify the contribution of old stars in the list of candidates.
We use optical photometric and spectroscopic data, as well 
as X-ray data from the \textit{ROSAT} satellite. We attempt to determine 
the nature of these stars using the information given by different 
age indicators such as the lithium line at 6707.8~\AA, 
the X-ray emission level, and the chromospheric activity. Isochrone fitting 
is used as well to place constraints on the age spread in the young 
moving groups. In particular, we explore the Local Association age spread
caused by the higher number of candidates in our sample.

\begin{table}
\caption{Young moving groups}                 
\label{tab:MGs}      
\centering                          
\begin{tabular}{l c c c}        
\hline\hline                 
\noalign{\smallskip}
Moving Group & Age [Myr] & Associated cluster(s) \\          
\noalign{\smallskip}
\hline                        
\noalign{\smallskip}
Local Association  &  $10 - 300$ & Pleiades, $\alpha$ Per, M34, \\
                   &             & $\delta$ Lyr, NGC 2516, IC 2602 \\
\noalign{\smallskip}
IC 2391 Supercluster & $80 - 250$ & IC 2391 \\
\noalign{\smallskip}
Castor             &    $200$    & $-$     \\
\noalign{\smallskip}
Ursa Major         &  $300 - 500$  & Ursa Major \\
\noalign{\smallskip}
Hyades Supercluster &   $650$    & Hyades, Praesepe \\
\noalign{\smallskip}
\noalign{\smallskip}
\hline                                   
\end{tabular}
\end{table}
%

    \begin{figure*}[!t]
    \centering
    \includegraphics[width=8.6cm]{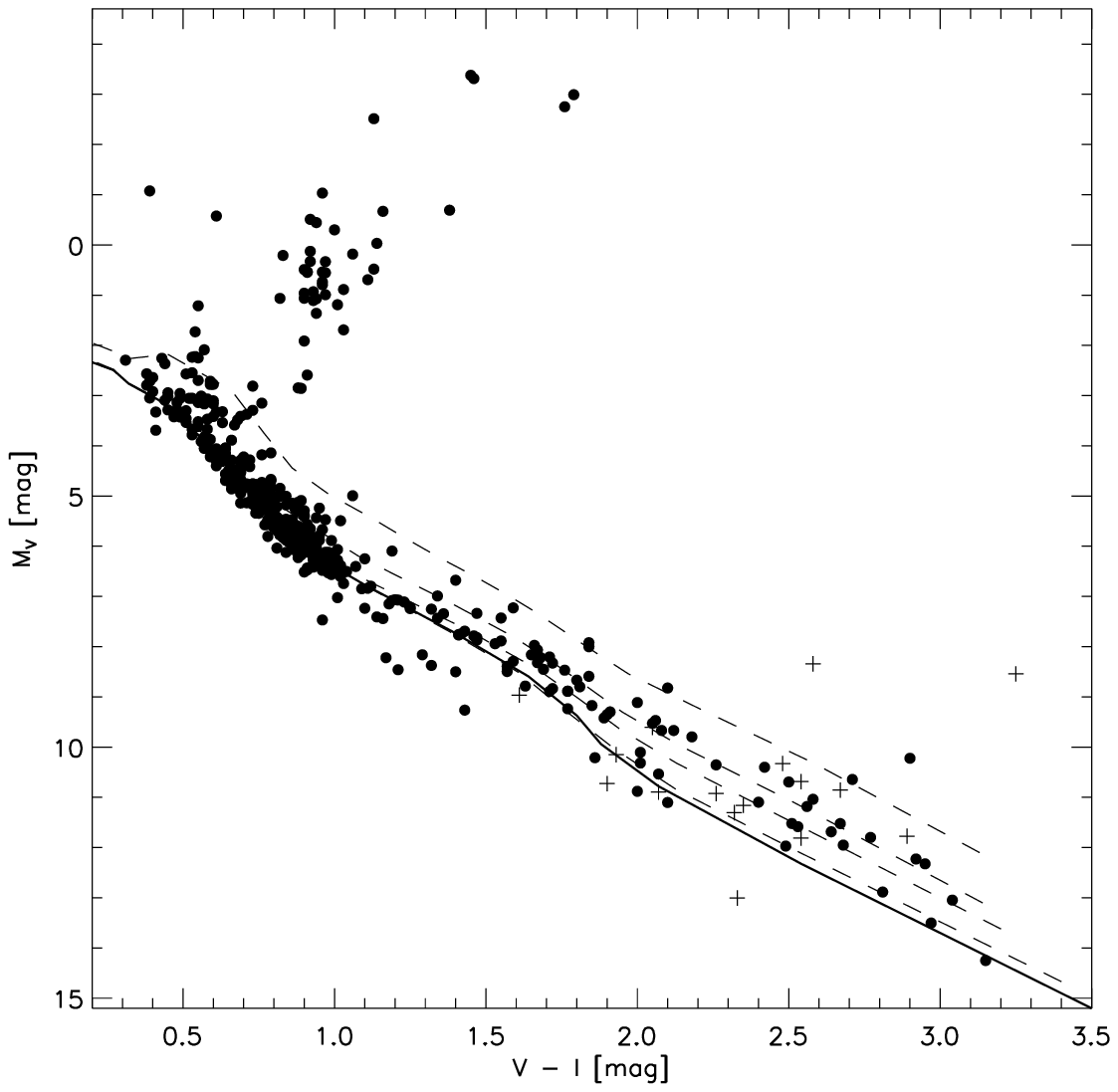}
    \includegraphics[width=8.6cm]{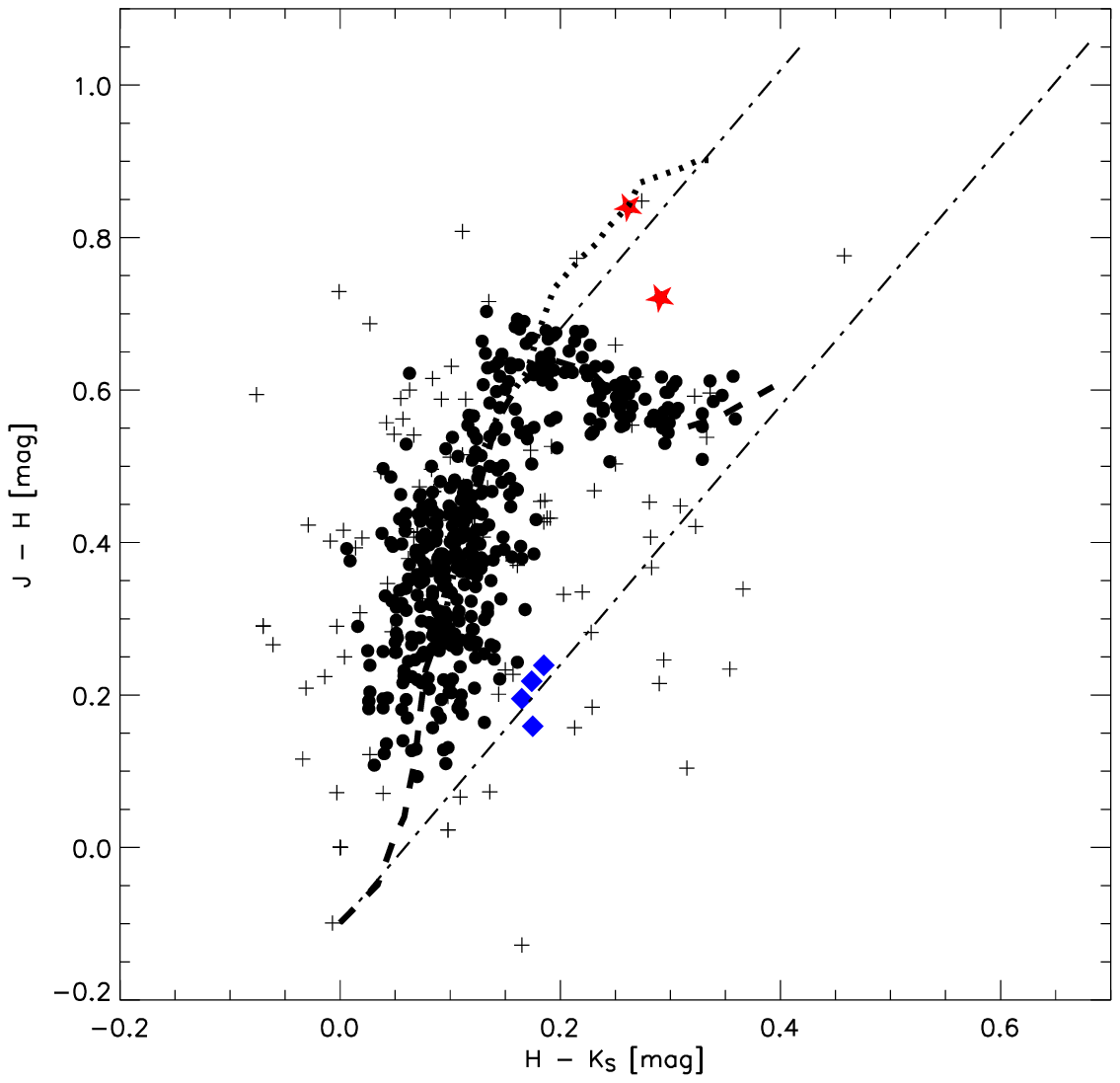}
    \caption{\textbf{Left:} $M_\mathrm{V}$ vs. $V-I$ of the sample of
             stars possible members of the young moving groups in 
             the Hipparcos catalogue. Filled
             circles are the stars with reliable photometry. Crosses are
             stars with uncertain $\sigma_{V-I} \ge 0.1$\,mag. Pre-main
             sequence isochrones of \citet{sie00} are plotted as dashed
             lines (from top to bottom: 10, 20, 50, and 120 Myr), while
             the continuous line represents the ZAMS. The dot
             situated over the 10\,Myr isochrone is AT\,Mic, which is 
             a visual binary. Hipparcos gives the visual magnitude of the 
             system. Assuming that both stars in the system have spectral
             type M4.5, the correction of $M_\mathrm{V}$ would be 
             $\sim 0.75$, situating AT Mic just on the 10~Myr isochrone, 
             which is coherent with its membership in the $\beta$~Pictoris
             moving group.
             \textbf{Right:} near-IR color--color diagram of the stars
             in our sample. Filled circles are stars with good quality
             2MASS photometry (quality flag $=$ 'AAA'). The star-like
             symbols represent FK\,Ser and HD\,142764. The filled grey 
             diamonds
             are stars with precise photometry situated out of the main
             sequence track. Stars with error-bars larger than 0.1 mag
             are plotted as crosses. Dashed and dotted lines are the
             main-sequence track and the giant branch defined by
             \citet{bes88}, transformed here to the 2MASS system. The
             dotted-dashed lines represent the reddening vector for
             A0 and M0 dwarfs.
            }
             \label{fig:opt_IR}
    \end{figure*}

\section{Data compilation}
\label{s:sample}

IN our study, we considered late-type stars proposed to be members of the 
young stellar kinematic groups of \citet{mon01a}. This sample includes only 
late-type (spectral types F--M) field stars in the solar neighborhood 
--\,typically at distances $d \le 100$\,pc\,-- selected by the authors
from several compilations of the literature \citep[see][for a complete 
bibliographical list and the selection criteria]{mon01a}.
Although
the sample is biased towards active stars because it was compiled mainly 
from studies of magnetic (chromospherical and coronal) activity 
and surveys of young late-type stars, many
stars were taken from works in which no differentiation was made between old 
and young stars, such as the catalog of the 100 nearby stellar systems given 
by the Research Consortium on Nearby Stars 
(RECONS\footnote{RECONS: {\tt http://www.recons.org/}}), or 
the search for kinematic groups in the solar neighborhood by \citet{orl95}.
After the compilation, \citet{mon01a} used only kinematical criteria to 
assign each star to a moving group.
A list of the young moving groups is given in Table~\ref{tab:MGs}, 
together with the stellar clusters historically associated with 
each group.
We also give the age -- or range of ages -- of each moving group 
given in the literature, i.e., the Local Association \citep{asi99}, 
Hyades supercluster \citep{sku99}, Ursa Major moving group \citep{sod93, asi99,
kin03}, IC 2391 supercluster \citep{egg91}, and Castor moving group 
\citep{bar98}. In general, the ages were determined by isochrone fitting
of the members of the moving groups, together with some spectroscopic criteria,
such as the lithium abundance and chromospheric activity.
%
To the initial sample of 535 stars, we added the 21 members of the moving 
groups from the spectroscopic survey of late-type stars in the solar 
neighborhood of \citet{lop05} and \citet{lop08} that were not included in 
\citet{mon01a}. Therefore, our sample contains a total of 556 late-type stars.
\citet{lop08} determined the lithium abundance, rotational and radial 
velocity, and level of magnetic activity of a sample of 144 late-type stars 
members of the moving groups, using data from echelle spectrographs. 
For these 144 stars, we applied different age indicators to our data from high 
resolution ($\Delta \lambda / \lambda \sim 50000$) optical observations.


Optical photometric information ($B_\mathrm{T}-V_\mathrm{T}$ and $V-I$ 
colors) was taken from the Tycho-2 and Hipparcos catalogues \citep{esa97, 
hog00}. The Tycho-2 colors ($B_\mathrm{T}-V_\mathrm{T}$) were 
used to determine $V$ magnitudes following the \mbox{Sect. 1.3} of 
the Hipparcos \textit{Introduction and Guide to the Data} \citep{esa97}. 
We note 
that the values of the $V-I$ colors in the Hipparcos catalogue are given in 
the Cousin photometric system. Of the 556 stars in the initial sample, only
12 do not have Tycho-2 entries, while 62 are not included in the Hipparcos 
catalogue. 
The information on the distance provided by Hipparcos was combined with 
the visual magnitude ($V$) of each star to determine its absolute visual 
magnitude $M_\mathrm{V}$. For the stars not included in the Hipparcos 
catalogue, we used the distances given in \citet{mon01a}, who either took 
them from the literature or determined spectroscopic parallaxes.
We searched for IR counterparts by cross-correlating our sample with
the 2MASS\footnote{The Two  Micron All Sky Survey (2MASS) 
is available at http://www.ipac.caltech.edu/2mass/} database.
We initially used a search radius of $r = 10$\,arcsec, although only 9 IR 
sources (2\% of the sample) were then found at $r > 5$\,arcsec. It is remarkable 
that no IR source was found for 6 stars with the chosen radius, 
all of which are faint dwarfs ($V > 11$\,mag). One hundred and six IR
counterparts show large errors in the 2MASS colors ($\sigma \ge 0.1$\,mag). 
This was taken into account when studying their position in the near-IR 
color--color diagram (see $\S$~\ref{s:phot}).


We also searched for X-ray counterparts of the stars 
in both the ROSAT All-Sky Survey Bright Source Catalogue (RASS-BSC) and the 
Faint Source Catalogue (RASS-FSC). A search radius of 30\,arcsec was adopted,
bearing in mind the ROSAT X-ray object coordinate determination accuracy.
A total of 341 stars (61\% of the moving-groups sample) are matched with 
X-ray sources, out of which about 6\% are expected to be spurious
\citep[see][for a detailed discussion]{gui98}. Since we are concerned with 
statistical studies, a contamination of 6\% ought not to affect any of the 
conclusions drawn in this paper. 
%
To determine the X-ray fluxes, we used the count rate-to-energy 
flux conversion factor ($CF$) relation found by \citet{sch95}:
\begin{center}
$CF = (A \cdot HR + B) \cdot 10^{-12}$ ergs cm$^{-2}$ counts$^{-1}$
\end{center}
where $HR$ is the hardness-ratio of the star in the ROSAT energy band 
0.1--2.4\,KeV, $A = 5.30$, and $B = 8.31$. We note that this $CF$ relation is 
valid for main-sequence stars. \citet{hun96} found that for late-type 
giants and super-giants, $B = 8.7$. X-ray fluxes were determined by
multiplying the $CF$ by the count-rate of the sources in the same 
band. Fluxes were later transformed into luminosity using the distances of the 
stars. 
Since the $CF$ and the count rate ($CR$) are defined for the ROSAT 
energy band 0.1--2.4\,keV, the X-ray luminosity $L_\mathrm{X}$ is also 
defined in this band.




\section{Evidences for the presence of an old population}
\label{s:phot}

In Fig.~\ref{fig:opt_IR} (left panel), we plot $M_\mathrm{V}$ versus $V-I$ for 
the possible members of the moving groups. Pre-main sequence (PMS) 
isochrones of 10, 30, 50, and 120 Myr of \citet{sie00} are overplotted as 
dashed lines, while the continuous line represents the ZAMS. 
Our stars are situated mainly on the main-sequence (MS) locus, but a large
quantity are located above the ZAMS. Evolved (giant) stars are 
distinctive because of their high luminosity. The results are 
compatible with the sample containing a mixture of 
MS, PMS and 
some evolved stars. The PMS population shows a range of ages of approximately 
$10-120$\,Myr, as deduced by comparing the data with the isochrones. 

Only one T\,Tauri star (FK\,Ser) is known to be present in the 
sample, but it is not plotted in the $M_\mathrm{V}$ versus $V-I$ diagram because 
we did not find any value of $V-I$ in the literature. In the near-infrared 
color--color diagram (Fig.~\ref{fig:opt_IR}, right panel), 
there are only two stars with precise 2MASS photometry, 
i.e., the already mentioned FK\,Ser and HD\,142764, a K5
dwarf with $A_V = 1.8$ mag \citep{eir01} that are situated above the MS.
The remaining stars are
on or close to the MS. Some stars with photometry of lower accuracy 
(plotted as crosses in 
the figure) are outside the MS track, but their position could be an artifact 
of the imprecise photometry. It is interesting to note that two of the 
four stars near the reddening vector (filled diamonds) host  
`hot jupiters', while the other two are known to be PMS stars. 

    \begin{figure}
    \centering
    \includegraphics[width=8.6cm]{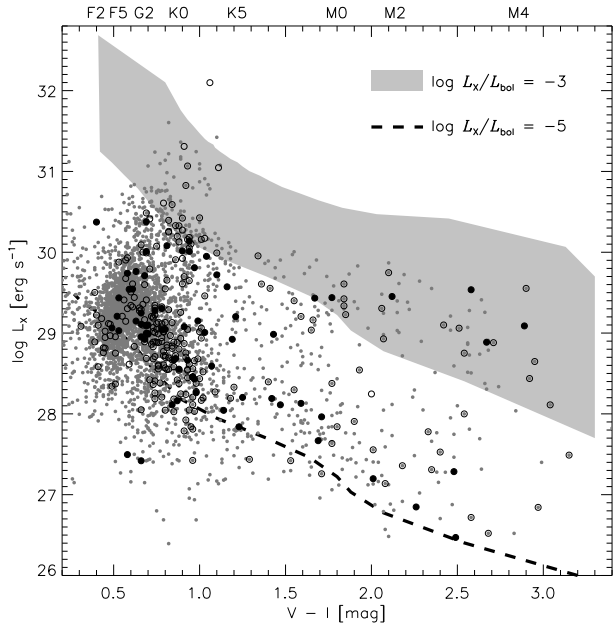}
    \caption{X-ray luminosity vs. $V-I$ color diagram for the stars in our
             sample (circles). Filled circles are the stars with 
             high-resolution spectroscopic observations. Small dots are the 
             stars from the Hipparcos catalogue with $d \le 100$ pc. The grey
             area represents the region of 
             $\log L_\mathrm{X}/L_\mathrm{bol} = -3$, with the lower edge
             corresponding with the ZAMS of the pre-main sequence models
             of \citet{sie00}, and the upper edge with the 1 Myr isochrone.
             The dashed line represents the $\log L_\mathrm{X}/L_\mathrm{bol}
             = -5$ relation.
            }
             \label{fig:Lx_VI}
    \end{figure}
In Fig.~\ref{fig:Lx_VI}, we plot $L_\mathrm{X}$ versus $V-I$ for the sources in 
our sample. For comparison, we also plot the stars in the Hipparcos catalogue 
at $d \le 100$~pc with a \textit{ROSAT} counterpart (small dots). 
In a similar way to \citet{zic05}, 
we observe a clear trend of $L_\mathrm{X}$ decreasing with decreasing stellar 
temperature in our sample, but we also observe 
that
the sources are
divided into two strips (or branches): the lower envelope of the first one 
coincides with the locus of the $\log L_\mathrm{X}/L_\mathrm{bol} = -5$ 
relation (dashed line); in contrast, the sources in the upper strip are 
situated in the region delimited by the $\log L_\mathrm{X}/L_\mathrm{bol}=-3$ 
relation (grey area), corresponding with the saturation value for active stars. 
A similar behavior was noticed by \citet{dae07} using 
other colors in a more restricted sample of field stars (see their Figure 2).
%
It has been observed that the gap in the $L_\mathrm{X}$ versus $V-I$ diagram is 
a consequence of the decrease in X-ray emission with age 
\citep[see again][for a complete discussion]{dae07}. The upper 
envelope of the $\log L_\mathrm{X}/L_\mathrm{bol}$ distribution is populated
by stars exhibiting lithium absorption lines, typical of young stars, while the 
lower branch is typically populated by older stars without lithium absorption 
line. 
The combination of both X-ray emission and lithium abundances allows us
to distinguish between old and young stars in the main sequence.

    \begin{figure}[!t]
    \centering
    \includegraphics[width=8.6cm]{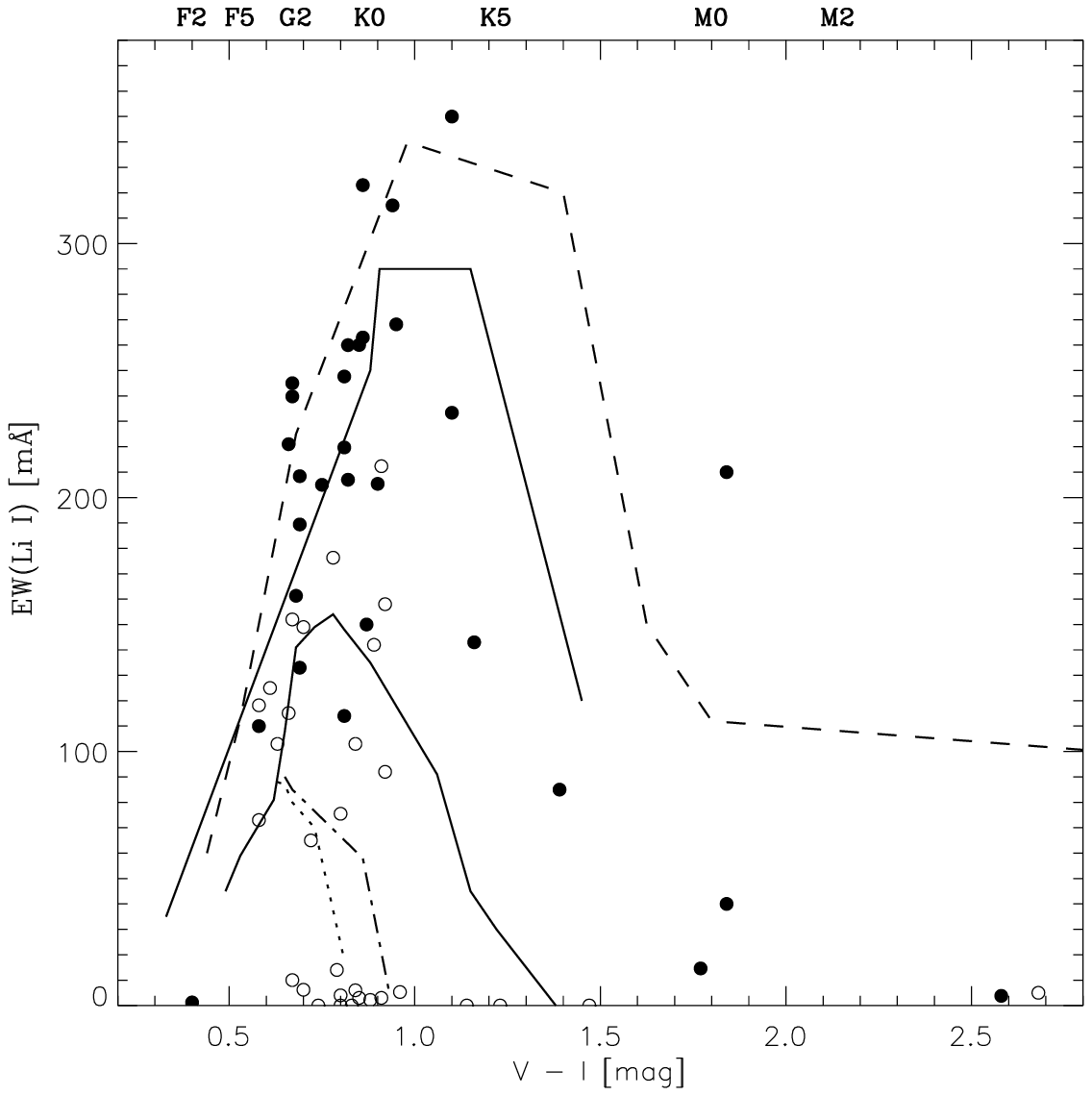}
    \caption{Lithium ($\lambda6707.8 \ \AA$ line) equivalent width of the
             stars of the Local Association. The upper envelopes of the
             stars of some stellar clusters are plotted: dashed line for IC~2391
             \citep[$10-35$ Myr;][]{mon01b}, continuous line for the Pleiades 
             \citep[$80-120$ Myr;][]{sod93b}, 
             doted-dashed line for the Coma Berenices cluster 
             \citep[$\sim 400$ Myr;][]{lop05}, 
             and dotted line for the Hyades \citep[$\sim 650$ Myr;][]{sod93b}. The lower envelope 
             of the Pleiades is also plotted as a continuous line. The filled circles 
             are stars with $\log L_\mathrm{X}/L_\mathrm{bol}~\ge~-4$.
            }
             \label{fig:LA_li}
    \end{figure}

    \begin{figure*}[!t]
    \centering
    \includegraphics[width=8.6cm]{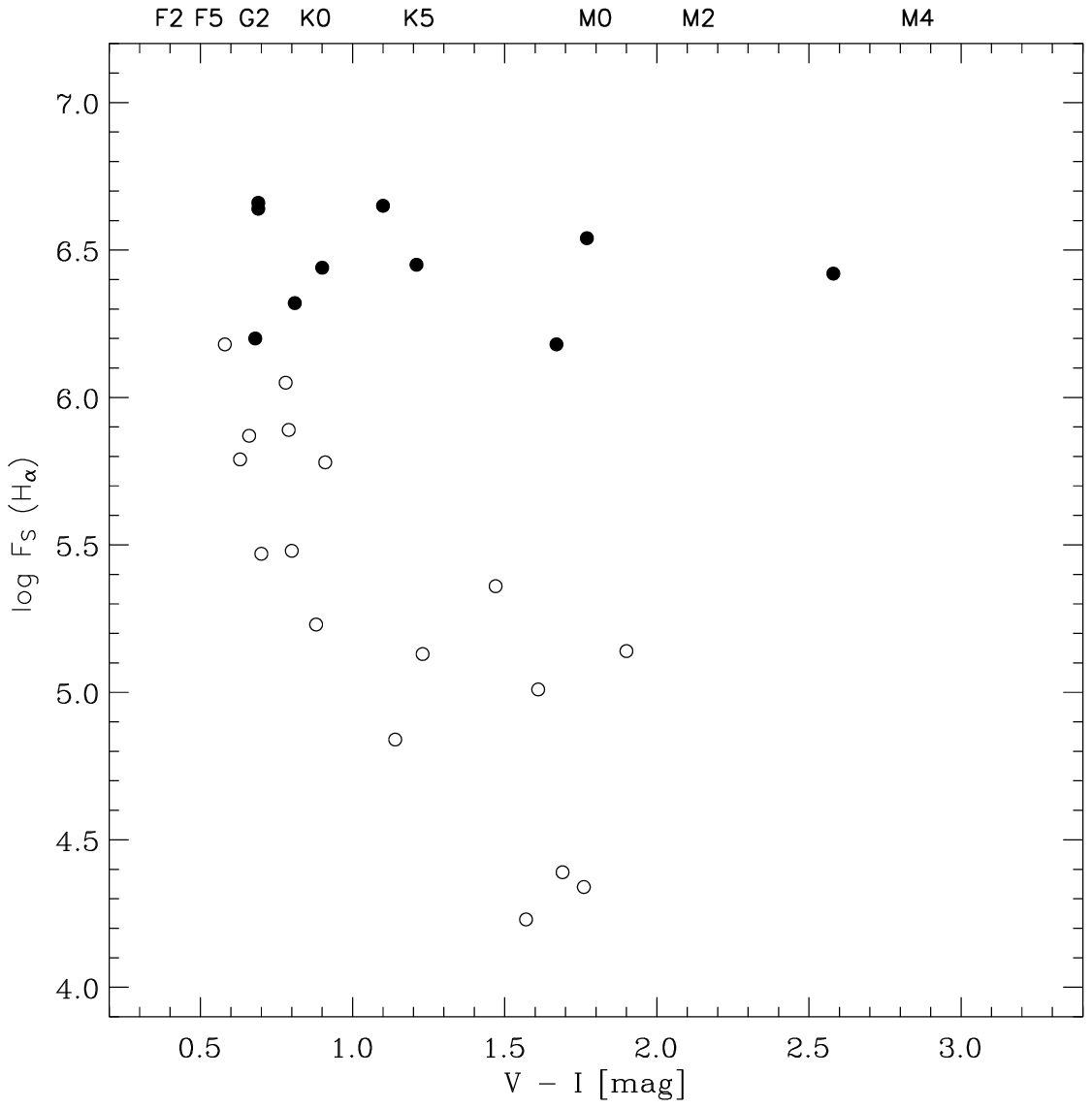}
    \includegraphics[width=8.6cm]{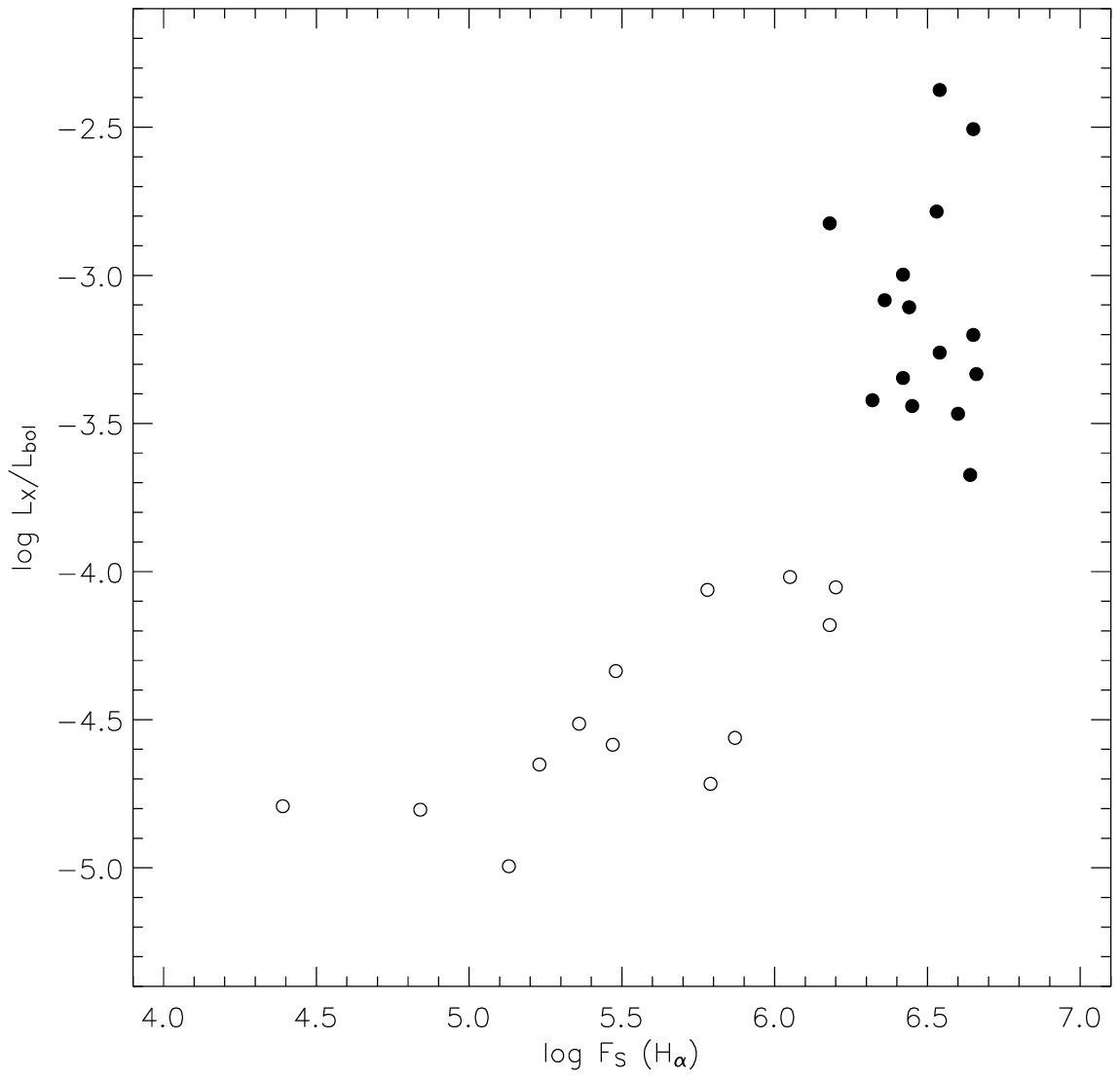}
    \caption{\textbf{Left:} $\log F_\mathrm{S}$(H$_{\alpha}$) vs. $V-I$
             of the members of the Local Association for which we have
             a measure of the equivalent width of H$_{\alpha}$. The filled 
             circles are stars with $\log L_\mathrm{X}/L_\mathrm{bol} \ge -4$.
             \textbf{Right:} $\log L_\mathrm{X}/L_\mathrm{bol}$ vs. 
             $\log F_\mathrm{S}$(H$_{\alpha}$) of the members of the Local 
             Association. The symbols are the same as in the left panel.
            }
             \label{fig:lxlbol_LA}
    \end{figure*}

\section{Old star population contamination in the Local Association}
\label{sec:Xray_LA}

Due to the large number of possible members in our sample, the Local Association
is the most suitable moving group for quantifying the fraction
of old stars among the candidate members of the moving groups. 
It is also the moving group that contains candidates 
with a higher spread in ages and differences in the stellar properties.
We used the information
compiled by us from the literature (see $\S$~\ref{s:sample}) to determine the physical
properties and, in particular, the age of each candidate.

In Fig.~\ref{fig:LA_li}, we plot the equivalent width of the lithium line $\lambda$6707.8~\AA\ 
versus the $V - I$ color of the stars considered to be members 
of the Local Association. For comparison, we also plot the upper envelopes of some 
clusters with accurately determined ages. From the figure, we can identify two populations:
the first one has equivalent widths similar to or even higher than those of 
the stars of the Pleiades, and the second one has equivalent widths that are below the Hyades upper 
envelope. Most of the stars in the first group exhibit also strong X-ray emission, the filled 
circles representing stars with $\log L_\mathrm{X}/L_\mathrm{bol} \ge -4$. They 
also have high levels of chromospheric activity. Figure~\ref{fig:lxlbol_LA} presents the 
flux in the H$_\alpha$ line in the candidates of the Local Association and its relation 
with the X-ray luminosity. 
Moreover, the stars with high 
X-ray emission levels exhibit a saturation in the H$_\alpha$ flux at each spectral type 
(Fig.~\ref{fig:lxlbol_LA}, left panel). All these results are compatible with the stars with
higher equivalent widths of the lithium line \ion{Li}{i} in our sample
being, at least, as young as the Pleiades (indicated by the upper and lower envelopes
in Fig.~\ref{fig:LA_li}). Although it is impossible to estimate their 
age, its range must be $10-120$~Myr.
The stars with equivalent widths of  \ion{Li}{i} below the lower envelope of 
the Pleiades show levels of chromospheric and X-ray 
emission typical of field stars. 
This group of stars is dominated by non-members of the moving group.

Assuming that all stars with low X-ray and chromospheric emission levels (and
low equivalent widths of lithium in FGK stars) are non-members of the 
Local Association, we 
find that the contamination of the initial sample of candidates is $\sim 50\%$. This
number, nevertheless, must be taken as an upper limit since our sample is incomplete.
In particular, the number of matches between our sample and the RASS decreases 
with increasing distance. At $d \le 40$ pc, approximately 70\% of our stars are 
cross-identified, while at $d \le 50$ pc, only
a half of them have a RASS counterpart. This suggests that some X-ray 
emitters at large distances are undetected because of the 
flux limit of the RASS, producing a bias in our sample. We cannot 
reject these stars \textit{a priori} as members of the moving groups. 
Using only the stars at $d \le 40$ pc, for which we are more complete, 
the contamination of our sample by non-members is $\sim 30\%$, which we 
propose to be a more reliable value. 
We note that our result is valid only for the candidates identified by
\citet{mon01a} and is closely related to the adopted selection criteria.
\citet{fam08} observed a higher percentage of field stars in a sample of K and M 
giant stars in the Pleiades Moving Group. \citet{ber06} also presented a detailed study 
of the Taurus-Auriga T association, 
where the authors identified 53\% of field stars among 217 possible 
pre-main sequence stars in this region.

\section{Contamination of the other moving groups}


We cannot apply the above analysis to the other moving groups since their 
number of candidates is too small to draw robust conclusions.
Furthermore, the other moving groups are, on average, older than 
the Local Association and the age indicators that we have used are
effective only for ages less than few hundreds million years. The lithium abundance, 
for instance, decays very rapidly for late-K and M dwarfs and may be 
barely detectable at the age of the Pleiades for late-type stars, while 
X-ray and chromospheric activity indicators may even be used at slightly
older ages. In agreement with their age, only the members of the 
Castor Moving Group show higher values of equivalent width of \ion{Li}{i}, but the 
number of stars for which we have data is not high enough to reach robust conclusions. 

\begin{figure}
    \centering
    \includegraphics[width=8.6cm]{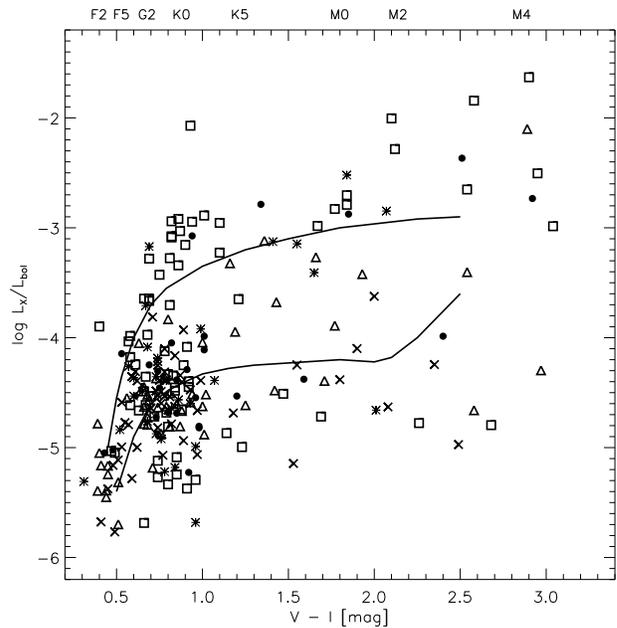}
    \caption{$\log L_\mathrm{X}/L_\mathrm{bol}$ vs. $V-I$ of the members of
             the Local Association (squares), the Castor MG (filled circles),
             Ursa Major MG (triangles), Hyades SC (x symbols), and
             IC 2391 SC (asterisks) . 
             The upper continuous line is the median of the Pleiades
             members. The continuous line at the bottom is the median of the 
             stars of the Hyades. 
            }
             \label{fig:lxlbol_Ca}
    \end{figure}

Studying the X-ray emission can still help us to separate stars of 
different age. In Fig.~\ref{fig:lxlbol_Ca}, 
we plot $\log L_\mathrm{X}/L_\mathrm{bol}$ versus $V-I$ for all stars of  
our sample with RASS counterparts. Each moving group is represented by 
a different symbol. 
The two continuous lines indicate the median X-ray emission of 
Pleiades and Hyades members determined by us from ROSAT data.
The high X-ray 
emitters of the Local Association exhibit X-ray emission levels above those of
the Pleiades members.
A similar behavior is observed in the Castor Moving Group. 
Both the Local Association and the Castor candidates seem to show a similar fraction 
of contaminating old field stars. The dK and dM of the Ursa Major Moving Group
($300 - 500$ Myr) have $\log L_\mathrm{X}/L_\mathrm{bol} \sim -4$, between the 
Castor Moving Group and the Hyades Supercluster, which have low 
$L_\mathrm{X}/L_\mathrm{bol}$ ratios. However,
for dF and dG stars, it is more difficult to identify this effect, because 
of the poor sensitivity X-ray luminosity to age. 
Finally, the late-type stars in the IC 2391 Supercluster show high X-ray emission. 
This would agree with the IC 2391 being as young as the Pleiades. However, the
number of candidates in this moving group is quite small and we cannot draw 
significant conclusions.
%
%

\section{Summary and conclusions}

We have attempted to quantify the contamination by old field stars among 
candidate members of the young stellar kinematic groups, or moving groups. Our sample 
was selected from possible members of the young stellar kinematic
groups studied by \citet{mon01a}. We compiled photometric, spectroscopic, and X-ray data 
from the literature. We cross-correlated the sample with the Tycho-2, 
Hipparcos, 2MASS, and RASS databases. Spectroscopic data were taken from 
\citet{lop05} and \citet{lop08}. 

The general properties of the stars for which we analyzed data indicate that there 
are probably two different age populations. We used the equivalent 
width of the lithium line $\lambda6708.8$~\AA\ and the X-ray luminosities to constrain
the ages of the candidates of the Local Association, which is the moving group 
containing the youngest stars. The analysis indicates that 
the contribution of old field stars to the sample of candidates of this moving group
is $\sim 50\%$ if we used the whole sample. A more accurate study using 
only those stars at $d \le 40$~pc, in which our sample of X-ray sources is 
nearly complete, showed that the contribution of non-members is, in fact, 
approximately 30\%. 
This number must be considered to represent the contamination by old main 
sequence stars of the \citet{mon01a} sample of candidate members of the 
young moving group. A higher percentage of interlopers among 
possible members of the Pleiades Moving Group was determined 
in \citet{fam08}.
The age of the other moving groups does not permit members to be
separated univocally from non-members, since the properties of the candidates 
are similar to those of field stars. However, we found that the X-ray data can 
be used to distinguish, at least, between stars of age $< 200$~Myr and older stars. 

Our result is consistent with that found for the Hyades Supercluster by 
\citet{fam07}. In that study, the authors concluded that the moving group consist 
of stars coeval with the Hyades cluster, which presumably formed 
in the same star formation episode, and old stars with similar space motion.
The samples of candidate members of the classical young moving groups 
($< 650$~Myr) contain a non-negligible amount of old ($> 1$~Gyr) field stars.
With regard to this matter, several studies suggest that the 
\textit{three arm-like shape} observed in the $UV$-plane is caused by a
non-axisymmetric potential \citep{fam05} and that the same non-axisymmetric
motion affecting old-field stars should also perturb the molecular clouds
where star formation occurs \citep{xu06}. This possibly explains
for the location of the young stellar associations such as TW Hya or $\eta Cha$ 
inside the Local Association in the $UV$-plane. 
In addition, this scenario could explain the problem of the apparent lack of post-T Tauri
stars close to star-forming regions \citep{her78, sod98}, which is inferred from the
increasing number of them identified in loose 
associations and moving groups \citep{bub07,tor08,mon09}.

We have extended our conclusions for the Local Association to the 
other young moving groups. The contamination by old stars in the moving groups  
should be considered in studies of the history of stellar birthrate in the solar 
neighborhood. Only by using data of different wavelengths, in particular X-ray and 
optical spectroscopic data, is it possible to distinguish between the two age 
populations. More detailed studies in this area will enable us to determine univocally
the contribution of young stars to the stellar population in the solar vicinity, to compare
with the predictions of Galactic models.

\begin{acknowledgements}
J. L\'opez-Santiago acknowledges financial support by the Marie Curie 
Fellowship contract No. MTKD-CT-2004-002769 and financial contribution by
MERG-CT-2007-046535. The Madrid group acknowledges financial contribution by
the Universidad Complutense de Madrid and the Programa Nacional de 
Astronom\'{\i}a y Astrof\'{\i}sica of the Spanish Ministerio de Educaci\'on y
Ciencia (MEC), under grants AYA2005--02750 and AYA2008-000695 and to the 
PRICIT project S-0505/ESP-0237 (ASTROCAM) of the Comunidad de Madrid. 
G. Micela acknowledges financial contribution from contract ASI--INAF I/088/06/0.
We thanks the referee for useful comments that allowed to improve 
our manuscript.
This publication makes use of data products from the Two Micron All Sky
Survey, which is a joint project of the University of Massachusetts and
the Infrared Processing and Analysis Center/California Institute of
Technology, funded by the National Aeronautics and Space Administration
and the National Science Foundation.
This research made use of the SIMBAD database, operated at the CDS,
Strasbourg, France.
\end{acknowledgements}

\end{document}